\documentclass[aps,prl,reprint]{revtex4-1}

\usepackage{amssymb}
\usepackage{mathtools}
\usepackage{amsmath}
\usepackage{amstext}
\usepackage{graphicx,epsfig}
\usepackage{enumitem}

\usepackage{tikz}

\definecolor{darkblue}{rgb}{0.15,0.35,0.55}
\definecolor{reddish}{rgb}{0.65, 0.2, 0.2}
\usepackage[linktocpage=true]{hyperref}
\hypersetup{
colorlinks=true,
citecolor=darkblue,
linkcolor=reddish,
urlcolor=darkblue,
pdfauthor={},
pdftitle={},
pdfsubject={}
}

\newcommand{\nn}{ \nonumber\\}
\newcommand{\Q}{\mathcal{Q}}
\newcommand{\Qvec}{\vec{\Q}}

\newcommand{\mvec}{\vec{\mu}}

\newcommand{\rd}{{\rm d}}

\begin{document}
\title{A Universal Relation Between Corrections to Entropy and Extremality}
\author{Garrett Goon}
\email{ggoon@andrew.cmu.edu}

\author{Riccardo Penco}
\email{rpenco@andrew.cmu.edu}
\affiliation{Department of Physics, Carnegie Mellon University, 5000 Forbes Ave,
Pittsburgh, PA 15217, USA}

\begin{abstract}
Perturbative corrections to General Relativity alter the expressions for both the entropy of black holes and their extremality bounds.   We prove a universal relation between the leading corrections to these quantities.  The derivation is purely thermodynamic and the result also applies beyond the realm of gravitational systems.  In scenarios where the correction to the entropy is positive, our result proves that the perturbations decrease the mass of extremal black holes, when holding all other extensive variables fixed in the comparison.  This implies that the extremality relations of a wide class of black holes display Weak Gravity Conjecture-like behavior.
\end{abstract}

\maketitle

\section{Introduction}

From a high energy physics perspective, General Relativity is a low-energy Effective Field Theory (EFT) of the gravitational sector. Its UV completions generically include additional degrees of freedom whose low-energy signatures are captured by higher-derivative corrections.  For macroscopically large black holes, these additional operators perturbatively alter the familiar relationships between the quantities which specify the black hole state and derived properties such as the Hawking temperature, entropy, and extremality bounds.

In this note, we derive a universal relation between the leading corrections to the extremality bound and entropy of generic black holes. Consider a gravitational theory permitting black holes characterized by mass $M$ and additional quantities $\Qvec$.  Associated to these solutions are an entropy $S_{0}(M,\Qvec)$ and temperature $T_{0}(M,\Qvec)$.  States of fixed $\Qvec$ generically have a minimum mass set by an extremality bound of the form $M>M_{\rm ext}^{0}(\Qvec)$.  

Now consider perturbatively changing the theory in a manner controlled by a parameter $\epsilon$, defined precisely later in \eqref{GenericFreeEnergy}.  The extremality bound and entropy then become $\epsilon$-dependent and are changed to the form $M>M_{\rm ext}(\Qvec,\epsilon)$ and $S(M,\Qvec,\epsilon)$, where $M_{\rm ext}(\Qvec,0)=M_{\rm ext}^{0}(\Qvec)$ and $S(M,\Qvec,0)=S_{0}(M,\Qvec)$.  In this note, we prove the following, exact result under mild assumptions:
\begin{align}
\mkern-9mu \frac{\partial M_{\rm ext}(\Qvec,\epsilon)}{\partial \epsilon}=\lim_{M\to M_{\rm ext}(\Qvec,\epsilon)}\mkern-9mu-T\left (\frac{\partial S(M,\Qvec,\epsilon)}{\partial \epsilon}\right )_{M,\Qvec}\ ,\label{ExactRelationIntroduction}
\end{align}
where $T=T(M,\Qvec,\epsilon)$ is the corrected temperature.

Using the above, we then derive further, approximate relations which hold at leading order in a perturbative expansion.  For instance, \eqref{ExactRelationIntroduction} implies a leading-order expression of the form
  \begin{align}
\Delta M_{\rm ext}(\Qvec)
&\approx-T_{0}(M,\Qvec)\Delta S(M,\Qvec )\Big|_{M\approx M^{0}_{\rm ext}(\Qvec)}\ ,\label{ApproxRelation2Introduction}
  \end{align}
where $\Delta S(M,\Qvec )$ and $\Delta M_{\rm ext}(\Qvec)$ are the leading corrections to the entropy of a state with fixed $M,\Qvec$ and to the extremality bound, respectively.  The meaning of $M\approx M^{0}_{\rm ext}(\Qvec)$ and the assumptions under which \eqref{ApproxRelation2Introduction} is valid are discussed in detail below. As thermodynamic quantities do not generically admit analytic expansions in $\epsilon$ near extremality, the conditions under which \eqref{ExactRelationIntroduction} follows from \eqref{ApproxRelation2Introduction} are not entirely trivial.

Our results are intimately connected to the Weak Gravity Conjecture (WGC) \cite{ArkaniHamed:2006dz}. One facet of the WGC is that the $U(1)$ charge-to-mass ratio  of extremally charged black holes should be larger than unity in any gravitational EFT which admits a consistent UV completion.   In particular, the WGC posits that black holes support $Q/M\ge 1$ at extremality for generic masses and the bound is expected to only approach the classical Einstein-Maxwell extremality result $Q/M=1$ in the $M\to \infty$ limit.    See \cite{Cheung:2014ega,Bellazzini:2019xts,Hamada:2018dde} for previous EFT-based arguments supporting this version of the conjecture.

The connection between \eqref{ApproxRelation2Introduction} and the WGC is due to the work of \cite{Cheung:2018cwt,Cheung:2019cwi}.  The authors demonstrate  that, under certain assumptions, higher-derivative corrections generate $\Delta S (M,\Qvec) >0$ for thermodynamically stable black holes. A similar idea was also suggested in \cite{Kats:2006xp}.  This line of reasoning was applied to the case of a single $U(1)$ charge $Q$ in \cite{Cheung:2018cwt}, where higher-derivative corrections to the entropy and extremality bound were computed explicitly and found to obey
\begin{align}
\Delta M_{\rm ext}(Q)\propto \Delta S(M,Q)\Big|_{M\approx Q}\ \label{CheungRemmenRelation}
\end{align}
with a negative constant of proportionality.  If $\Delta S>0$, then the single $U(1)$ form of the WGC follows immediately.

The result \eqref{ApproxRelation2Introduction} is a generalization of \eqref{CheungRemmenRelation} and it follows that a generic perturbed extremal black hole is less massive than its unperturbed counterpart with the same quantum numbers, if $\Delta S>0$. This statement is the generalization of having an increased $Q/M$ ratio in the case of multiple non-trivial quantum numbers.  Hence, if the $\Delta S>0$ conjecture of \cite{Cheung:2018cwt} is correct, \eqref{ApproxRelation2Introduction} proves that the extremality curves of a wide class of black hole systems display WGC-like behavior.  Conversely, assuming the WGC is correct, \eqref{ApproxRelation2Introduction} demonstrates that the conjecture also implies a preferred sign for shifts to the entropy of near extremal black holes.

We have confirmed that \eqref{ApproxRelation2Introduction} is consistent with the results in \cite{Cheung:2018cwt,Cheung:2019cwi,Reall:2019sah,Cano:2019oma,Cano:2019ycn} and in the final section of this note we explicitly verify \eqref{ExactRelationIntroduction} and \eqref{ApproxRelation2Introduction} in the case of charged, asymptotically anti-de Sitter, four-dimensional black holes.

The derivation of \eqref{ExactRelationIntroduction} is purely thermodynamic and is not specific to black holes.  Relatedly, the proof has no explicit dependence on such details as the spacetime dimensionality or the matter content of the theory. In a general context, \eqref{ExactRelationIntroduction} is a relation between the change in the energy of a system at zero temperature and the change in the entropy at fixed extensive variables due to an alteration of the underlying dynamics.    While \eqref{ExactRelationIntroduction} and its approximate descendants may therefore find interesting applications beyond the regime of black hole physics, in this note we remain focused on gravitational systems.

\noindent \textbf{Conventions}: We work in Euclidean signature and our curvature conventions are $R^{\rho}{}_{\sigma\mu\nu}=\partial_{\mu}\Gamma^{\rho}_{\nu\sigma}+\ldots$ and $R_{\sigma\nu}=R^{\rho}{}_{\sigma\rho\nu}$.  Natural units are used throughout: $G_{N}=k_{B}=\hbar=c=1$.

\noindent \textbf{Note Added}: The central derivation in this letter has been improved. The original letter only contained the approximate relation \eqref{ApproxRelation2Introduction} which was derived for $M\approx M_{\rm ext}^{0}(\Qvec)$, whereas the current version proves the exact result \eqref{ExactRelationIntroduction}, from which \eqref{ApproxRelation2Introduction} follows.  The relation \eqref{ExactRelationIntroduction} explains subtleties which can arise in the strict $M=M_{\rm ext}^{0}(\Qvec)$ limit and is consistent with  \cite{Loges:2019jzs,Cano:2019oma,Cano:2019ycn} which probe this regime.  References \cite{Cano:2019oma,Cano:2019ycn} appeared after the initial version of this letter.

\section{General Argument}

The argument for \eqref{ExactRelationIntroduction} follows from considering the effects of perturbative corrections to the free energy of generic thermodynamic systems.  This problem was addressed in \cite{Reall:2019sah} for the case of spinning, four-dimensional black holes.  

Consider a thermodynamic system characterized by the entropy $S$ and the collection of additional extensive variables $\Qvec$ such that the energy of the system is $M(S,\Qvec)$.  The first law is then of the form
\begin{align}
  \rd M=T\rd S+\mvec\cdot\rd \Qvec\ , \label{FirstLaw}
  \end{align}
  with $\mvec$ a set of generalized chemical potentials.  In a black hole context, $\Qvec$ may contain conserved quantities, such as various angular momenta and $U(1)$ charges, as well terms not associated to any conservation law, like the ``volume" of the black hole \cite{Kastor:2009wy} and the value of moduli fields at infinity \cite{Gibbons:1996af} in scenarios where variations of these latter parameters are considered.
  
  Let us work in the ensemble where all of the quantum numbers $\Qvec$ are allowed to fluctuate.  The following argument also generalizes to cases where an arbitrary subsets of $\Qvec$ fluctuates; we only focus on the present ensemble for clarity of presentation.  The associated thermodynamic potential $G(T,\mvec)$ arises from Legendre transforming $M(S,\Qvec)$ over $S$ and $\Qvec$ in the usual manner. We will simply refer to $G$ as the ``free energy" in the following.

We assume that the above free energy consists of a dominant, zeroth order piece, $G_{0}(T,\mvec)$, plus a perturbative correction $\Delta G(T,\mvec)$.  In this note, unperturbed quantities will carry a $0$ subscript or superscript and will often be referred to as being ``classical."  Following \cite{Reall:2019sah}, we insert a convenient counting parameter $\epsilon$ in front of $\Delta G$ such that $G(T,\mvec)\longrightarrow G(T,\mvec,\epsilon)$ where
\begin{align}
G(T,\mvec,\epsilon)&\equiv G_{0}(T,\mvec)+\epsilon\Delta G(T,\mvec)\ .\label{GenericFreeEnergy}
\end{align}
The quantities $S$, $\Qvec$, and $M$ are given by:
\begin{align}
S(T,\mvec,\epsilon)&=-\left (\frac{\partial G}{\partial T}\right )_{\mvec,\epsilon}\nn
\Q_{i}(T,\mvec,\epsilon)&=-\left (\frac{\partial G}{\partial \mu_{i}}\right )_{T,\mu_{j\neq i},\epsilon}\nn
M(T,\mvec,\epsilon)&= G(T,\mvec,\epsilon)+TS+\mvec\cdot\Qvec\ , 
\label{BasicThermoRelations}
\end{align}
which are the standard thermodynamic relations.

When the microscopic dynamics of the system are described by an action of the form $I=I_{0}+\epsilon\Delta I$, with $I_{0}$ the leading term and $\Delta I$ a perturbative correction, $\Delta G$ is directly determined by $\Delta I$, at leading order in $\epsilon$.  In the specific case of black holes, $G(T,\mvec)$ is given by~\cite{Gibbons:1976ue}
\begin{align}
\beta\, G(T,\mvec)&=I[g_{\mu\nu}(T,\mvec),\ldots]\ ,\label{EuclideanAction}
\end{align} 
where $\beta=1/T$ is the inverse Hawking temperature, $I$ is the Euclidean action, $g_{\mu\nu}=g_{\mu\nu}(T,\mvec)$ is the Euclideanized metric solution parameterized in terms of $T$ and $\mvec$, the ellipses represent other possible fields, and the corresponding entropy $S$ in \eqref{BasicThermoRelations} is the Wald entropy \cite{Wald:1993nt}.  A subtraction prescription is typically required to render \eqref{EuclideanAction} finite \cite{Gibbons:1976ue,Emparan:1999pm,Balasubramanian:1999re}. For gravitational systems, the unperturbed action $I_{0}$ might be Einstein-Hilbert, Einstein-Maxwell, etc.~and $\Delta I$ could represent higher-derivative corrections, extra matter fields, or lower derivative terms such as a cosmological constant, for example. We only require that the effects of $\Delta I$ are appropriately small.
The Euclidean solution is also corrected due to the change in the theory, such that $g_{\mu\nu}=g^{0}_{\mu\nu}+\epsilon\Delta g_{\mu\nu}$, with $g^{0}_{\mu\nu}$ a solution of $I_{0}$.  The terms in \eqref{GenericFreeEnergy} are, at leading order,
 \begin{align}
 \beta G_{0}&=I_{0}[g^{0}_{\mu\nu},\ldots]\nn
  \epsilon\beta \Delta G&=\epsilon\Delta I[g^{0}_{\mu\nu},\ldots]\nn
  &\quad +I_{0}[g^{0}_{\mu\nu}+\epsilon\Delta g_{\mu\nu},\ldots]-I_{0}[g^{0}_{\mu\nu},\ldots]\ .\label{G0DeltaGForGravitationalSystems}
 \end{align}
 Up to $\mathcal{O}(\epsilon)$, the final line above can only receive a contribution from boundary terms and is expected to vanish if these (and any necessary subtraction prescriptions) are chosen to respect the proper variational principle, essentially by definition, in which case $\Delta g_{\mu\nu}$ is not required and $\Delta G\propto\Delta I$.  This fact was explicitly demonstrated in \cite{Reall:2019sah}, for the case of asymptotically flat spacetimes and also occurs in the example of the following section.

We can derive a universal relation from \eqref{GenericFreeEnergy} by assuming that we can invert \eqref{BasicThermoRelations} to switch amongst the variables $T$, $\mvec$, $M$, and $\Qvec$ and that the perturbative corrections to the entropy respect the condition
 \begin{align}
	\lim_{T\to 0} T \left (\frac{\partial S(T,\Qvec,\epsilon)}{\partial \epsilon}\right )_{T,\Qvec} = 0 \ , \label{ThirdLawAssumption}
\end{align}
which can be interpreted as a version of the third law.

We begin by considering the perturbed extremality bound, which is now $M>M_{\rm ext}(\Qvec,\epsilon)$ where
\begin{align}
M_{\rm ext}(\Qvec,\epsilon)&\equiv \lim_{T\to 0}M(T,\Qvec,\epsilon)\ .
  \end{align}  
  Let us characterize the effect of the perturbative corrections by computing the $\epsilon$-derivative of $M$ at fixed $T,\Qvec$.  Using \eqref{BasicThermoRelations} and the chain rule, one finds
  \begin{align}
  \left (\frac{\partial M}{\partial \epsilon}\right )_{T,\Qvec}&=\left (\frac{\partial}{\partial\epsilon}\left (G+TS+\mvec\cdot\Qvec\right )\right )_{T,\Qvec}\nn
  &=\left (\frac{\partial  G }{\partial \mvec}\right )_{T,\epsilon}\! \!\cdot\left (\frac{\partial  \mvec }{\partial \epsilon}\right )_{T,\Qvec}
  +\left (\frac{\partial  G }{\partial \epsilon}\right )_{T,\mvec}\nn
  &\quad
  +T\left (\frac{\partial  S }{\partial \epsilon}\right )_{T,\Qvec}
  +\Qvec\cdot\left (\frac{\partial  \mvec }{\partial \epsilon}\right )_{T,\Qvec}\ ,
  \end{align}
  where we considered $G$ as a function of $(T,\mvec,\epsilon)$ while $S$ and $\mvec$ were treated as functions of $(T,\Qvec,\epsilon)$.  Due to \eqref{BasicThermoRelations}, the first and final terms above cancel and we are left with
  \begin{align}
  \left (\frac{\partial M}{\partial \epsilon}\right )_{T,\Qvec}=\Delta G(T,\mvec(T,\Qvec,\epsilon))+T\left (\frac{\partial  S }{\partial \epsilon}\right )_{T,\Qvec}\ ,\label{dMdepsilon}
  \end{align}
   where we used that $\left (\frac{\partial G}{\partial \epsilon}\right )_{T,\mvec}=\Delta G(T,\mvec)$, by definition. Finally, if we take the $T\to 0$ limit and use the third law assumption \eqref{ThirdLawAssumption}, we find
     \begin{align}
  \lim_{T\to 0}\left (\frac{\partial M}{\partial \epsilon}\right )_{T,\Qvec}=\lim_{T\to 0}\Delta G(T,\mvec(T,\Qvec,\epsilon))\ .\label{dMdepsilonT0}
  \end{align}

   Next, the change in the entropy at fixed $M,\Qvec$ is characterized by $\left (\frac{\partial S}{\partial \epsilon}\right )_{M,\Qvec}$.  A straightforward generalization of the argument in \cite{Reall:2019sah} (see \cite{Loges:2019jzs}, also) yields the result
   \begin{align}
    -T\left (\frac{\partial S}{\partial \epsilon}\right )_{M,\Qvec}&=\Delta G\left (T(M,\Qvec,\epsilon),\mvec(M,\Qvec,\epsilon)\right )\ .\label{dSdepsilon}
    \end{align}  The proof is similar to the derivation of \eqref{dMdepsilon}: it is a straightforward exercise in the chain rule and the use of thermodynamic identities, including the first law \eqref{FirstLaw}.

 It is then evident that \eqref{dMdepsilonT0} and \eqref{dSdepsilon} coincide if the latter is evaluated at the extremal point $M=M_{\rm ext}(\Qvec,\epsilon)$ corresponding to $T=0$ in the perturbed theory:
 \begin{align}
\mkern-10mu  \lim_{T\to 0}\left (\frac{\partial M}{\partial \epsilon}\right )_{T,\Qvec}\mkern-7mu &=\mkern-7mu \lim_{M\to M_{\rm ext}(\Qvec,\epsilon)}\mkern-20mu -T(M,\Qvec,\epsilon)\left (\frac{\partial S}{\partial \epsilon}\right )_{M,\Qvec} .\label{ExactRelation}
 \end{align}
 As argued below, we expect the left side of \eqref{ExactRelation} to be finite, in which case $\left (\partial S/\partial\epsilon\right )_{M,\Qvec}$ must diverge as $\sim T^{-1}$ as $M\to M_{\rm ext}(\Qvec,\epsilon)$, though $S(M,\Qvec,\epsilon)$ itself need not diverge in this limit. The above is an exact identity for any system of the form \eqref{GenericFreeEnergy}, irrespective of whether $\Delta G$ is small in any sense.  However, we anticipate that in applications the role of $\epsilon$ will be played by a combination of EFT coefficients and the free energy will only take the form \eqref{GenericFreeEnergy} in a leading-order expansion, in which case \eqref{ExactRelation} is exact only for the truncated system. 
 
The result \eqref{ExactRelation} relates properties of corrected states in the perturbed theory.  Through suitable approximations, it is also possible to interpret \eqref{ExactRelation} as a comparison between states in the classical and corrected theories, but there are subtleties in this analysis.  In particular, the right side of the relation is suggestive of a comparison of entropies of states with fixed $M,\Qvec$, but cannot immediately be understood as such if $M_{\rm ext}(\Qvec,\epsilon)<M_{\rm ext}^{0}(\Qvec)$.  This is due to the fact that the derivative is evaluated at $M_{\rm ext}(\Qvec,\epsilon)$ and there was no state carrying mass $M_{\rm ext}(\Qvec,\epsilon)$ and quantum numbers $\Qvec$ in the unperturbed theory, in this scenario. 
 
 In the aforementioned case, we can instead evaluate right side of \eqref{ExactRelation} at the unperturbed extremal point $M_{\rm ext}^{0}(\Qvec)$, rather than $M_{\rm ext}(\Qvec,\epsilon)$.  This corresponds to a valid comparison between states in the classical and perturbed theories, as such a state exists in both cases, by assumption. The result is the approximate relation
 \begin{align}
\mkern-10mu  \lim_{T\to 0}\left (\frac{\partial M}{\partial \epsilon}\right )_{T,\Qvec}\mkern-7mu &\approx\mkern-7mu \lim_{M\to M_{\rm ext}^{0}(\Qvec)}\mkern-20mu -T(M,\Qvec,\epsilon)\left (\frac{\partial S}{\partial \epsilon}\right )_{M,\Qvec} \ ,\label{ApproxRelation1}
 \end{align}
 with the error controlled by $\epsilon$. 
 
 In this regime, the temperature  must also be proportional to a power of $\epsilon$.  Moreover, one expects the power to be fractional, in the generic case, due to the fact that the unperturbed extremal state will typically only continue to exist in the corrected theory for one sign of $\epsilon$ (an imaginary temperature indicating the erasure of the state).  Similar consideration apply to $S(M,\Qvec,\epsilon)$.  For instance, possible expansions for $S$ and $T$ which are consistent with results in the literature \cite{Cano:2019ycn,Loges:2019jzs} are
 \begin{align}
  S(M_{\rm ext}^{0}(\Qvec),\Qvec,\epsilon)&\approx S_{\rm ext}^{0}(\Qvec)+\epsilon^{\lambda}\Delta S(M_{\rm ext}^{0}(\Qvec),\Qvec)\nn
  T(M_{\rm ext}^{0}(\Qvec),\Qvec,\epsilon)&\approx \epsilon^{1-\lambda}\Delta T(M_{\rm ext}^{0}(\Qvec),\Qvec)\ ,\label{TSExpansionNonanalytic}
 \end{align}
 where $0<\lambda<1$.   In contrast, we expect the mass to have an analytic expansion in $\epsilon$ about $T=0$,
  \begin{align}
 \lim_{T\to 0}M(T,\Qvec,\epsilon)&\approx M_{\rm ext}^{0}(\Qvec)+\epsilon \Delta M_{\rm ext}(\Qvec)\ ,\label{MEpsilonExpansion}
 \end{align}
 since, on physical grounds, zero temperature states of fixed $\Qvec$ should exist in both the classical and unperturbed theories for generic perturbations, and hence for either sign of $\epsilon$, unlike those of fixed $(M,\Qvec)$ with $M$ near extremality.  Plugging \eqref{TSExpansionNonanalytic} and \eqref{MEpsilonExpansion} into \eqref{ApproxRelation1}, we are then left with the leading-order result
  \begin{align}
 \Delta M_{\rm ext}(\Qvec)\! &=-\lambda\Delta T(M,\Qvec)\Delta S(M,\Qvec)\Big|_{M=M_{\rm ext}^{0}(\Qvec)}\ ,\label{ApproxRelation1Example}
 \end{align}
 in this example. Behavior such as \eqref{TSExpansionNonanalytic} where the temperature is dominated by the $\epsilon$-dependent corrections was interpreted in \cite{Cheung:2018cwt} as a breakdown of the perturbative calculation, but \eqref{ApproxRelation1} remains valid in this regime as long as the higher-order corrections in $\epsilon$ are small.
 
 Finally, by further modifying \eqref{ExactRelation} we can derive an approximate relation which is applicable regardless of whether $M_{\rm ext}(\Qvec,\epsilon)$ is larger or smaller than $M_{\rm ext}^{0}(\Qvec)$.  Specifically, we evaluate the right side of \eqref{ExactRelation} a mass range slightly above  $M_{\rm ext}^{0}(\Qvec)$ where corrections to the temperature are additionally small, i.e., where we satisfy
\begin{align}
\mkern-8mu\frac{M-M_{\rm ext}^{0}(\Qvec)}{M_{\rm ext}^{0}(\Qvec)}\ll 1\ , \frac{T(M,\Qvec,\epsilon)-T_{0}(M,\Qvec)}{T_{0}(M,\Qvec)}\ll 1\ .\label{RequirementsGenericBound}
\end{align}
 The latter condition further indicates that $M>M_{\rm ext}(\Qvec,\epsilon)$, meaning that states in this mass range exist in the perturbed theory for either sign of $\epsilon$.  Hence, we expect that both $T$ and $S$ will admit analytic expansions in $\epsilon$ of the form
 \begin{align}
 T(M,\Qvec,\epsilon)&\approx T_{0}(M,\Qvec)+\epsilon \Delta T(M,\Qvec)\nn
 S(M,\Qvec,\epsilon)&\approx S_{0}(M,\Qvec)+\epsilon \Delta S(M,\Qvec)\ .\label{TSExpansionAnalytic}
 \end{align}
 with $T_{0}\gg \epsilon\Delta T$, $S_{0}\gg \epsilon\Delta S$.    In practice, we access this region by working at $M=M_{\rm ext}^{0}(\Qvec)\times (1+\delta)$ with $\delta \ll 1$, as required by the first half of \eqref{RequirementsGenericBound}, and with $\delta$ generically bounded non-trivially from below by the second half of \eqref{RequirementsGenericBound}.  As in \eqref{ApproxRelation2Introduction}, we refer to this part of parameter space as $M\approx M_{\rm ext}^{0}(\Qvec)$ and evaluating the right side of \eqref{ExactRelation} here gives
  \begin{align}
\mkern-10mu  \lim_{T\to 0}\left (\frac{\partial M}{\partial \epsilon}\right )_{T,\Qvec}\mkern-15mu &\approx\mkern-3mu  -T(M,\Qvec,\epsilon)\left (\frac{\partial S}{\partial \epsilon}\right )_{M,\Qvec}\Big|_{M\approx M_{\rm ext}^{0}(\Qvec)}\mkern-10mu \label{ApproxRelation2}
 \end{align} with the error now controlled both by $\epsilon$ and $\delta$.  Inserting \eqref{TSExpansionAnalytic} into \eqref{ApproxRelation2} results in the claimed, leading-order relation \eqref{ApproxRelation2Introduction}.

 We conclude this section by commenting on related work.  The relation \eqref{ApproxRelation2Introduction} valid for $M\approx M_{\rm ext}^{0}(\Qvec)$ has been confirmed to be in agreement with the results in \cite{Cheung:2018cwt,Cheung:2019cwi,Reall:2019sah,Cano:2019ycn}.  A near-horizon metric based explanation for why the shifts to the black hole extremality bound and entropy are related was given in Sec.~6.3 of \cite{Cheung:2018cwt}, which gives a complementary argument for \eqref{ApproxRelation2Introduction} in the restricted case of gravitational systems.  The relation valid in the strict $M=M_{\rm ext}(\Qvec)$ limit \eqref{ApproxRelation1} has been checked against \cite{Loges:2019jzs,Cano:2019oma,Cano:2019ycn} and, again, exact agreement is found.  In particular, \cite{Loges:2019jzs,Cano:2019ycn} have results of the form \eqref{ApproxRelation1Example} with $\lambda=1/2$ while in \cite{Cano:2019oma} it was demonstrated that there are cases where \eqref{ApproxRelation1} trivializes with $\Delta M_{\rm ext}=T=0$, but $\Delta S\neq 0$.  We expect such degenerate cases to be highly non-generic.

 \section{Example: Charged Black Holes in ${\bf AdS_{4}}$}
 
 In this section, we explicitly confirm the relations \eqref{ExactRelationIntroduction} and \eqref{ApproxRelation2Introduction} for the case of extremally charged black holes in a four-dimensional, asymptotically anti-de Sitter (\textit{AdS}) spacetime.  First, we treat the scenario where cosmological constant plays the role of the perturbing parameter, in which case we confirm the exact result \eqref{ExactRelationIntroduction}.  Then, additional higher-derivative operators are included in the action and we verify the approximate $M\approx M_{\rm ext}(\Qvec)$ result \eqref{ApproxRelation2Introduction} for large, extremal black holes.
 
 The classical Euclidean action we consider is
\begin{align}
I_{0}[g,A]&=-\frac{1}{16\pi}\int_{\mathcal{M}}\rd^{4}x\,\sqrt{g}\, \left (R-F^{2}+\frac{6}{\ell^{2}}\right )\nn
&\quad-\frac{1}{8\pi}\int_{\partial\mathcal{M}}\rd^{3}x\, \sqrt{h}\, K\ ,\label{AdSEMAction}
\end{align} 
with $\mathcal{M}$ the spacetime manifold and $\partial\mathcal{M}$ its boundary.
Above, $F^{2}=F_{\mu\nu}F^{\mu\nu}$, $\ell$ is the AdS radius, $h_{ij}$ is the metric induced on $\partial \mathcal{M}$, and $K$ is the trace of extrinsic curvature of $\partial\mathcal{M}$.

The action \eqref{AdSEMAction} is divergent when evaluated on a solution and we will regulate these divergence as in \cite{Gibbons:2004ai} by subtracting off the action of empty \textit{AdS}.  The divergence could alternatively be treated using holographic counterterms along the lines of \cite{Emparan:1999pm,Balasubramanian:1999re}, for instance, but we do not pursue this direction here.

The black hole solution of \eqref{AdSEMAction} with charge $Q$ and mass $M$ has a background metric and vector potential, $g^{0}_{\mu\nu}\rd x^{\mu}\rd x^{\nu}\equiv\rd s^{2}_{0}$ and $A^{0}_{\mu}$, given by:
\begin{align}
\rd s_{0}^{2}&=\Sigma(r)\rd t^{2}+\Sigma(r)^{-1}\rd r^{2} +r^{2}\rd\Omega^{2}_{2}\nn
\Sigma(r)&\equiv 1-\frac{2M}{r}+\frac{Q^{2}}{r^{2}}+\frac{r^{2}}{\ell^{2}}\nn
A^{0}_{\mu}\rd x^{\mu}&=\frac{iQ}{r}\rd t\ ,\label{BackgroundSoln}
\end{align}
where $\rd\Omega^{2}_{2}$ is the standard line element on a two-sphere. There are, of course, more general solutions with non-trivial angular momentum and magnetic charge; see \cite{Caldarelli:1999xj}, for instance, for a study of \textit{AdS} Kerr-Newman thermodynamics in a more general setting.

We begin by confirming the exact result \eqref{ExactRelationIntroduction} for the above system, with $1/\ell^{2}$ playing the role of the perturbing parameter.  In order to make this explicit, we rescale $\ell\to \ell/\sqrt{\epsilon}$ in the following.  It is a standard exercise to derive the thermodynamic quantities associated to \eqref{BackgroundSoln} and the resulting mass and temperature of the system can be expressed via the exact relations
\begin{align}
M(S,Q,\epsilon)&=\frac{S+\pi Q^{2}}{2\sqrt{\pi S}}+\frac{\epsilon S^{3/2}}{2\pi^{3/2}\ell^{2}}\nn
T(S,Q,\epsilon)&=\frac{S-\pi Q^{2}}{4\sqrt{\pi}S^{3/2}}+\frac{3\epsilon\sqrt{S}}{4\pi^{3/2}\ell^{2}}\ .\label{AdSRNRelations}
\end{align}
The fact that the mass can naturally be written in the exact form $M=M_{0}(S,Q)+\epsilon \Delta M(S,Q)$ implies that \eqref{ExactRelationIntroduction} will in fact hold to all orders in $\epsilon$ for this system, as follows from the microcanonical ensemble version of the proof from the preceding section.  That is, we do not need to restrict to small black hole states which are only perturbatively corrected by $\ell$.   

A derivative of \eqref{AdSRNRelations} determines $\left (\partial S/\partial \epsilon\right )_{M,Q}$ to be
\begin{align}
-T\left (\frac{\partial S}{\partial\epsilon}\right )_{M,Q}&=\frac{S^{3/2}}{2\pi^{3/2}\ell^{2}}\ .
\end{align}
The extremal value of $S$ is found by solving \eqref{AdSRNRelations} at $T=0$. Plugging the result into the above gives the extremal value of the derivative, denoted here as:
\begin{align}
-T\left (\frac{\partial S}{\partial\epsilon}\right )_{M,Q}\Big|_{\rm ext.}&=\frac{\ell\left (-1+\sqrt{1+\frac{12 Q^{2}\epsilon}{\ell^{2}}}\right )^{3/2}}{12\sqrt{6}\, \epsilon^{3/2}}\ .\label{NegTdSAdSRNAtExtremality}
\end{align}
On the other hand, inserting this value of $S$ into the first relation in \eqref{AdSRNRelations} gives the extremality bound:
\begin{align}
M_{\rm ext}(Q,\epsilon)&=\frac{12Q^{2}+\ell^{2}\left (-1+\sqrt{1+12Q^{2}\epsilon/\ell^{2}}\right )}{3\sqrt{6\epsilon}\,\ell\sqrt{-1+\sqrt{1+12Q^{2}\epsilon/\ell^{2}}}}\ .\label{AdSRNClassicalExtremality}
\end{align}
Taking the derivative of the \eqref{AdSRNClassicalExtremality} and simplifying demonstrates that $\left (\partial M_{\rm ext}/\partial \epsilon\right )_{Q}$ precisely coincides with \eqref{NegTdSAdSRNAtExtremality}, confirming the exact relation \eqref{ExactRelationIntroduction}.

Next, we verify the approximate $M\approx M_{\rm ext}(\Qvec)$ relation \eqref{ApproxRelation2Introduction} by adding the following higher-derivative operators to the system:
\begin{align}
\Delta I&=\frac{\ell^{2}}{16\pi}\int_{\mathcal{M}}\rd^{4}x\, \sqrt{g}\Big (\alpha_{1}F^{4}\!+\!\alpha_{2} \ell^{2}F^{6}\!+\!\alpha_{3}\ell^{4}F^{8}\Big)\ ,\label{HigherDerivativeAction}
\end{align}
where the factors of $\ell$ were introduced for convenience. If tracking powers of $\epsilon$, each $\alpha_{i}$ would be accompanied by an $\epsilon$, but we omit these factors in what follows.  We have focused on a small set of higher-derivative matter operators chosen such that that empty \textit{AdS} with radius $\ell$ remains a solution of the corrected action and such that no additional boundary terms are required to render the action well-posed, choices made for simplicity.  Analyses involving curvature-dependent operators whose results are compatible with our relations can be found in \cite{Cheung:2018cwt,Cheung:2019cwi,Reall:2019sah,Cano:2019oma,Cano:2019ycn}, for instance.

The operators in \eqref{HigherDerivativeAction} change both the metric and vector potential, but neither correction is needed for the verification of \eqref{ApproxRelation2Introduction}.  The correction to the vector potential produced by \eqref{HigherDerivativeAction} changes the action by a boundary term which vanishes as $\partial\mathcal{M}$ is taken to infinity.  The correction to the metric also only changes the action by a boundary term and after regulating the result as described above, the result is again trivial as $\partial \mathcal{M}$ is taken to infinity, in direct analogy to what was found in \cite{Reall:2019sah}.

The free energy corresponding to the classical action \eqref{AdSEMAction} in the small-$T$ limit is
\begin{align}
G_{0}(T,\mu)&=-\frac{\ell\left (-1+\mu^{2}\right )^{3/2}}{\sqrt{3}}+\mathcal{O}(T\ell)\ ,\label{G0AdS}
\end{align} 
as determined by a standard calculation. The leading correction to the free-energy then arises from simply evaluating \eqref{HigherDerivativeAction} on the zeroth order solution \eqref{BackgroundSoln} expressed as a function of $\mu$ and $T$ using the zeroth order relations between $(M,Q)$ and $(\mu,T)$.  The calculation is straightforward and gives
\begin{align}
\Delta G(T,\mu)&=\alpha_{1}\frac{\sqrt{3}\ell\mu^{4}}{5\sqrt{\mu^{2}-1}}-\alpha_{2}\frac{2\mu^{6}\ell}{\sqrt{3}(\mu^{2}-1)^{3/2}}\nn
&\quad+\alpha_{3}\frac{36\sqrt{3}\ell\mu^{8}}{13(\mu^{2}-1)^{5/2}}+\mathcal{O}(T\ell)\ .\label{DeltaGAdS}
\end{align}

We use \eqref{G0AdS} and \eqref{DeltaGAdS} to study large, $Q\gg \ell$ extremal black holes for the remainder of this letter, with the limit taken for simplicity.  In the classical theory \eqref{AdSEMAction}, the extremality bound in this limit is \eqref{AdSRNClassicalExtremality}
\begin{align}
M^{0}_{\rm ext}(Q)=\frac{2Q^{3/2}}{27^{1/4}\ell^{1/2}}\left (1+\mathcal{O}\left (\frac{\ell}{Q}\right )\right )\ .\label{MextLargeAdS}
\end{align}
In order to find $\Delta M_{\rm ext}(Q)$, we evaluate $\Delta G(T,\mu)$ \eqref{DeltaGAdS} on the classical expression for $\mu(Q)$ in the $T\to 0$ limit, with the result
\begin{align}
\Delta M_{\rm ext}(Q)&=\lim_{T\to 0}\Delta G(T,\mu_0(Q,T))\nn
&=\frac{3^{1/4}Q^{3/2}}{65\ell^{1/2}}\Big (39\alpha_{1}-130\alpha_{2}+540\alpha_{3}\Big )\ ,\label{DeltaMLargeAdS}
\end{align}
valid at leading order in a $\ell/Q$ expansion.  By computing the free energy to higher orders in $T$, one can also determine $T(M,Q)$ near extremality.  Evaluating the result at $M=\frac{2Q^{3/2}}{27^{1/4}\ell^{1/2}}\times(1+\delta)$, the leading-order result is
\begin{align}
 T\approx\frac{12^{1/4}}{\pi}\frac{Q^{1/2}}{ \ell^{3/2}}\sqrt{\delta\!-\!\frac{3}{130}\left (39\alpha_{1}\!-\!130\alpha_{2}\!+\!540\alpha_{3}\right )}\ ,\label{TLargeAdS}
\end{align}
which is consistent with \eqref{DeltaMLargeAdS}. The result \eqref{TLargeAdS} also demonstrates that satisfying the requirements in \eqref{RequirementsGenericBound} demands
\begin{align}
1\gg\delta \gg \alpha_{1},\alpha_{2},\alpha_{3}\ .\label{LargeAdSdeltaRange}
\end{align}
There is no difficulty in satisfying the above for any natural values of the $\alpha_{i}$.  

Finally, we compute $-T(M,Q)\Delta S(M,Q)$  by again retaining higher order terms in the free energy and using \eqref{BasicThermoRelations}.  This straightforward exercise gives
\begin{align}
-T\Delta S\approx  \frac{3^{1/4}Q^{3/2}}{65\ell^{1/2}}\Big (39\alpha_{1}-130\alpha_{2}+540\alpha_{3}\Big )\ ,\label{NegTdSLargeAdS}
\end{align}
when evaluated in the range \eqref{LargeAdSdeltaRange}, up to relative $\mathcal{O}(\delta^{1/2}, \alpha_{i}/\delta,\ell/Q)$ corrections.  Comparing \eqref{DeltaMLargeAdS} and \eqref{NegTdSLargeAdS}, we see that \eqref{ApproxRelation2Introduction} is confirmed. We have also verified that the system defined by \eqref{G0AdS} and \eqref{DeltaGAdS} obeys the third law condition \eqref{ThirdLawAssumption} at leading order in $\alpha_{i}$.

\textbf{Acknowledgements}: We thank Cliff Cheung, Junyu Liu, Harvey Reall, Grant Remmen, Ira Rothstein, and  Jorge Santos for helpful discussions.  We also thank Brando Bellazzini, Yuta Hamada, Toshifumi Noumi, and Gary Shiu for comments made on the original version of this letter after its initial appearance.  The Mathematica package \texttt{xAct} \cite{[][``xAct: Efficient tensor computer algebra for the Wolfram Language" www.xact.es]xAct} was used extensively in the course of this work. The work of GG and RP is  partially supported by the National Science Foundation under Grant No. PHY-1915611.

\bibliographystyle{apsrev4-1}
\bibliography{Bibliography}

\end{document}